\documentstyle[prd,aps,preprint,tighten]{revtex}
\newcommand{\RR}{\hbox{$I$\kern-3.8pt $R$}}
\newcommand{\rr}{\hbox{$\scriptstyle I$\kern-2.4pt $\scriptstyle R$}}
\newcommand{\be}{\begin{equation}}
\newcommand{\ee}{\end{equation}}
\newcommand{\bea}{\begin{eqnarray}}
\newcommand{\eea}{\end{eqnarray}}

\newcommand{\sss}{\scriptscriptstyle}
\newcommand{\mn}{{\mu\nu}}
\newcommand{\starF}{\hbox{${}^{\rm *}$\kern-1.5pt $F$}}
\newcommand{\starFtilde}{\hbox{${}^{\rm *}$\kern-1.5pt $\tilde F$}}
\newcommand{\MAtilde}{\hbox{${}^{\scriptscriptstyle M}$
     \kern-7.0pt ${\tilde A}$}}
\newcommand{\EAtilde}{\hbox{${}^{\scriptscriptstyle E}$
     \kern-7.0pt ${\tilde A}$}}
\newcommand{\EAtildebold}{\hbox{${}^{\scriptscriptstyle E}$
     \kern-7.0pt ${\tilde{\bbox{A}}}$}}
\newcommand{\EAbar}{\hbox{${}^{\scriptscriptstyle E}$
     \kern-7.0pt ${\bar A}$}}
\newcommand{\M}{{\cal M}}
\renewcommand{\S}{{\cal S}}
\renewcommand{\H}{{\cal H}}
\newcommand{\E}{{\cal E}}

\newcommand{\G}{{\cal G}}
\begin{document}
\preprint{gr-qc/9412018} %places preprint number at
%%                         top right of title page
%\draft %omission stifles PACS numbers
\title{Black Hole Pair Creation and the Entropy Factor}
\author{J. David Brown}
\address{Department of Physics and Department of Mathematics,\\
   North Carolina State University, Raleigh, NC 27695--8202}
%\date{\today}
\maketitle
%%%%%%%%%%%%%%%%%%%%%%%%%%%%%%%%%%%%%%%%%%%%%%%%%%%%%%%%%%%%%%%%%%%%%%
\begin{abstract}
It is shown that in the instanton approximation the rate of creation
of black holes is always enhanced by a factor of the exponential of
the black hole entropy relative to the rate of creation of compact matter
distributions (stars). This result holds for any generally covariant
theory of gravitational and matter fields that can be expressed in
Hamiltonian form. It generalizes the
result obtained previously for the pair creation of magnetically
charged black holes by a magnetic field in Einstein--Maxwell theory.
The particular example of pair creation of electrically charged
black holes by an electric field in Einstein--Maxwell theory is
discussed in detail.
\end{abstract}
\pacs{???}
%%%%%%%%%%%%%%%%%%%%%%%%%%%%%%%%%%%%%%%%%%%%%%%%%%%%%%%%%%%%%%%%%%%%%%%%%
%%%%%%%%%%%%%%%%%%%%    CHAPTER ONE    %%%%%%%%%%%%%%%%%%%%%%%%%%%%%%%%%%
%%%%%%%%%%%%%%%%%%%%%%%%%%%%%%%%%%%%%%%%%%%%%%%%%%%%%%%%%%%%%%%%%%%%%%%%%
\section{Introduction}
In a recent analysis of the pair creation of magnetically charged black
holes by a magnetic field in Einstein--Maxwell theory, it was shown that
the creation rate is enhanced by a factor of $\exp(\S_{\sss BH})$,
where $\S_{\sss BH}$ is the black hole entropy, relative to the pair
creation rate for GUT monopoles \cite{GGS}. This result is important
because it provides a clue to the problem of the origin of black hole
entropy. In particular, it is consistent with the view that black holes
have $\exp(\S_{\sss BH})$ internal or horizon quantum states.

In this article the pair creation of non--extreme black holes (with
horizons identified) is considered in a general setting. The result is
always the same---the pair creation rate is enhanced by a factor of
$\exp(\S_{\sss BH})$ relative to the creation rate for a pair of
compact matter configurations (stars). This result holds for {\it any\/}
generally covariant theory of gravitational and matter fields that can be
expressed in Hamiltonian form. The enhancement in the black hole creation
rate is derived solely from the formal mathematical framework in which
the pair creation rate and the density of quantum states are expressed
as path integrals.

The enhancement in the black hole creation rate applies, in particular,
to the creation of electrically charged black holes
by an electric field in Einstein--Maxwell theory.  This result has
been anticipated \cite{GS,DGKT}:
Since the creation rate for magnetically charged black holes is enhanced by
the factor $\exp(\S_{\sss BH})$, by duality of the electromagnetic field
the creation rate for electrically charged black holes should be enhanced
by the same factor. Although this argument is correct physically,
the details of the
calculation for the electric case are not entirely obvious. The
apparent difficulty stems from the use of instanton
methods in which the leading order approximation to the creation
rate is related to the action of a classical solution, the
instanton. For the case of magnetically charged black holes and
magnetic fields \cite{GGS,GS,DGKT,Gibbons,Ross,DGGH,Yi,HHR}, the
instanton is obtained by the familiar substitution of $-it$ for $t$ in
the magnetic Ernst solution. The resulting instanton consists of a real
Euclidean metric and a real electromagnetic vector potential. On the
other hand, for the case of electrically charged black holes and electric
fields, substitution of $-it$ for $t$ in the electric Ernst solution
yields an instanton that consists of a real Euclidean metric and an
{\it imaginary\/} electromagnetic scalar potential. As shown here,
this result is correct and leads to the expected pair creation rate for
electrically charged black holes.

The appearance of an imaginary scalar potential is
familiar from the path integral construction of the partition function
for an electrically charged black hole \cite{BBWY,CPW}. If the black hole
is rotating, the shift vector for the instanton is imaginary as well
\cite{BMY,BY,CECS}. In general, instantons are stationary solutions
with the following properties: all fields that appear in the Hamiltonian
as Lagrange multipliers are imaginary, and the canonical
variables are real. These are the essential properties
that allow the instanton solution to match the corresponding
Lorentzian solution along a stationary surface.

The pair creation of electrically charged black holes in Einstein--Maxwell
theory is treated as a concrete example in this article. Thus, I begin in
Sec.~2 with a discussion of the electric Ernst solution and its
relationship,
through electromagnetic duality, to the magnetic Ernst solution. Section 3
contains a discussion of the connection between a general stationary
Lorentzian solution of the Einstein--Maxwell equations of motion and its
associated instanton. The instanton for the electric Ernst solution
is displayed explicitly. In Sec.~4, the pair creation rate for
electrically charged black holes in Einstein--Maxwell theory is
computed relative to the pair creation rate
for electrically charged stars. The analysis is generalized in Sec.~5 to
apply to any generally covariant theory of gravitational and matter fields.
The generalization requires a careful comparison of the formal path integral
derivations of black hole pair creation and black hole entropy.

It should be emphasized that the results of this paper rely on the
instanton approximation. Thus, the existence is assumed of instanton
solutions that describe the creation of black holes and compact matter
distributions, in the appropriate physical contexts. On the other hand,
it is not necessary that these classical solutions be known. Typically
they are not. (An exception is the Ernst solution in Einstein--Maxwell
theory.) The central result---the enhancement of black hole pair creation
by the factor $\exp(\S_{\sss BH})$---does not depend on the details of
the theory or the details of the instanton solutions.

%%%%%%%%%%%%%%%%%%%%%%%%%%%%%%%%%%%%%%%%%%%%%%%%%%%%%%%%%%%%%%%%%%%%%%%%
%%%%%%%%%%%%%%%%%%%%    CHAPTER TWO    %%%%%%%%%%%%%%%%%%%%%%%%%%%%%%%%%
%%%%%%%%%%%%%%%%%%%%%%%%%%%%%%%%%%%%%%%%%%%%%%%%%%%%%%%%%%%%%%%%%%%%%%%%
\section{Duality and the Electric Ernst Solution}
Let $\epsilon_{\mn\rho\sigma}$ denote the totally antisymmetric
tensor (volume element) with $\epsilon_{0123} = \sqrt{-g}$. The
dual of the electromagnetic field $F_{\mn}$ is defined by
\be
\starF^{\mn} = (1/2) \epsilon^{\mn\rho\sigma} F_{\rho\sigma}
\Longleftrightarrow F^{\mn} = -(1/2) \epsilon^{\mn\rho\sigma}
\starF_{\rho\sigma} \ .\eqnum{1}
\ee
The electric and magnetic fields are
\be
E^\mu = F^{\mn} U_\nu \ ,\qquad B^\mu = -\starF^{\mn} U_\nu
\ , \eqnum{2}
\ee
respectively, where $U^\mu$ is the unit normal vector field of a
family of spacelike
hypersurfaces. The electromagnetic stress tensor can be
written either in terms of $F_{\mn}$ or $\starF_{\mn}$:
\bea
4\pi T_\mn & = & F_{\mu\alpha} {F_\nu}^{\alpha} -
(1/4) g_\mn F_{\alpha\beta}F^{\alpha\beta} \eqnum{3a}\\
& = & \starF_{\mu\alpha} {\starF_\nu}^{\alpha} -
(1/4) g_\mn \starF_{\alpha\beta}\starF^{\alpha\beta}
\ .\eqnum{3b}
\eea
The classical equations of motion are the Einstein equations
$G_{\mn} = 8\pi T_{\mn}$ (with Newton's constant equal to 1)
and the Maxwell equations $d{\bbox{F}} = 0$ and
$d{\hbox{${}^{\rm *}$\kern-1.5pt ${\bbox{F}}$}} = 0$. The
electromagnetic field $F_\mn$ and its dual $\starF_\mn$ play a
symmetric role. If \{$g_\mn$, $F_\mn$\} = \{${\tilde g}_\mn$,
${\tilde F}_\mn$\} is a solution of the Einstein--Maxwell equations,
then \{$g_\mn$, $F_\mn$\} = \{${\tilde g}_\mn$, $-\starFtilde_\mn$\}
is also a solution. Here, the symbols ${\tilde g}_\mn$ and
${\tilde F}_\mn$ refer to specific tensors. Thus,
\{$g_\mn$, $F_\mn$\} = \{${\tilde g}_\mn$, ${\tilde F}_\mn$\}
implies that the electromagnetic field $F_\mn$ is given by the
tensor ${\tilde F}_\mn$, whereas \{$g_\mn$, $F_\mn$\} =
\{${\tilde g}_\mn$, $-\starFtilde_\mn$\} implies that the
electromagnetic field $F_\mn$ is given by the tensor
$-\starFtilde_\mn$ (which is minus the dual of ${\tilde F}_\mn$).
According to the definitions (2), the electric and magnetic fields
for the solution \{${\tilde g}_\mn$, $-\starFtilde_\mn$\} are
${\tilde B}^\mu$ and $-{\tilde E}^\mu$, respectively, where
${\tilde E}^\mu$ and ${\tilde B}^\mu$ are the electric and magnetic
fields for the solution \{${\tilde g}_\mn$, ${\tilde F}_\mn$\}.

The Ernst solution \cite{Ernst} describes a pair of oppositely
charged black holes accelerating apart in an electric or
magnetic field. The electric and magnetic cases are related
by duality as described above. In both cases the metric for the
spacetime region containing one black hole is
\be
{\tilde g}_{\mn}dx^\mu dx^\nu =\frac{\Lambda^2}{A^2(x-y)^2}
\Bigl[ G(y) dt^2 - G^{-1}(y) dy^2 + G^{-1}(x) dx^2 \Bigr]
+ \frac{G(x)}{A^2(x-y)^2\Lambda^2} d\varphi^2 \ ,\eqnum{4}
\ee
where
\bea
G(\xi) & = & (1+r_- A\xi)(1-\xi^2 - r_+ A\xi^3) \ ,\eqnum{5a}\\
\Lambda(x,y) & = & (1 + Bqx/2)^2 + \frac{B^2 G(x)}{4A^2(x-y)^2}
\ ,\eqnum{5b}
\eea
and $q^2 = r_+ r_-$. For the magnetic case, the electromagnetic
field is ${\tilde F}_\mn = \partial_\mu\MAtilde_\nu -
\partial_\nu\MAtilde_\mu$ where
\be
{\MAtilde}_\varphi = -\frac{2}{B\Lambda} (1 + Bqx/2) + k
\ .\eqnum{6}
\ee
For the electric case, the electromagnetic field is
$-\starFtilde_\mn = \partial_\mu\EAtilde_\nu -
\partial_\nu\EAtilde_\mu$ where
\bea
{\EAtilde}_t & = & - \frac{B}{2A^2} \biggl[ \frac{G(y)}{(x-y)^2}
(1 + Bqx - Bqy/2) + (1+r_- Ay)(1+r_+ Ay) (1-Bqy/2)\biggr]
\nonumber\\
& & + qy + k \ .\eqnum{7}
\eea
The magnetic Ernst solution is \{${\tilde g}_\mn$, $\MAtilde_\mu$\}
and the electric Ernst solution is \{${\tilde g}_\mn$, $\EAtilde_\mu$\}.

For both the electric and magnetic Ernst solutions, certain
restrictions must be imposed on the parameters $r_-$, $r_+$,
$A$, and $B$ \cite{GGS,GS,DGKT,Gibbons,Ross,DGGH,Yi,HHR}.
In particular, assume $r_+ A < 2/(3\sqrt{3})$ so that
the three roots of the cubic factor in $G(\xi)$ are real. For
non--extreme black holes, the smallest of these roots, $\xi_2$,
obeys $\xi_2 > -1/(r_-A)$. The angular coordinate $x$ is restricted
to $\xi_3 \leq x \leq \xi_4$, where $\xi_3$ and $\xi_4$ are the
two larger roots of the cubic factor in $G(\xi)$. The poles
$x=\xi_3$ and $x =\xi_4$ are free from conical singularities
if $G'(\xi_3) \Lambda(\xi_3)^2 = -G'(\xi_4) \Lambda(\xi_4)^2$
and the period in $\varphi$ is $4\pi \Lambda(\xi_3)^2/G'(\xi_3)$.
(Note, $\Lambda(\xi_3) = \Lambda(\xi_3,y)$ is independent of
$y$, and similarly for $\Lambda(\xi_4)$.) The black hole event
horizon is the null surface $y=\xi_2$, and the acceleration
horizon is the null surface $y=\xi_3$.

For the electric Ernst
solution the magnitude of the electric field on the
axis $x=\xi_3$ at spatial infinity ($y\to\xi_3$) is
$B G'(\xi_3)/(2 \Lambda(\xi_3)^{3/2})$. The magnitude of the electric
charge of the black hole is
\be
\frac{1}{4\pi} \oint {\hbox{${}^{\rm *}$\kern-1.5pt $d\EAtildebold$}}
= \frac{q(\xi_4 - \xi_3) \Lambda(\xi_3)^{3/2}}
{G'(\xi_3) (1 + Bq\xi_4/2)} \ .\eqnum{8}
\ee
The electric Ernst solution coincides with the electric Melvin
solution \cite{Melvin} at spatial infinity, and also in the limit of
vanishing $r_-$ and $r_+$. The metric for the electric Melvin solution
is the same as that for the magnetic Melvin solution, while the
electromagnetic field is determined by the scalar potential
${\EAtilde}_t = B z$. (The notation is that of
Ref.~\cite{GGS}, so here $B$ is the value of the
electric field on the $z$--axis).

%%%%%%%%%%%%%%%%%%%%%%%%%%%%%%%%%%%%%%%%%%%%%%%%%%%%%%%%%%%%%%%%%%%%%%%%
%%%%%%%%%%%%%%%%%%%%    CHAPTER THREE    %%%%%%%%%%%%%%%%%%%%%%%%%%%%%%%
%%%%%%%%%%%%%%%%%%%%%%%%%%%%%%%%%%%%%%%%%%%%%%%%%%%%%%%%%%%%%%%%%%%%%%%%
\section{Instanton Solutions}
The instanton that enters the calculation of the creation rate for
electrically charged black holes is obtained by the substitution
$t\to -it$ in the electric Ernst solution \{${\tilde g}_\mn$,
$\EAtilde_\mu$\}. It is useful to adopt a general notation and to
consider this step from a Hamiltonian point of view.
First, recall that the metric tensor and electromagnetic potential
can be split in space and time according to
\bea
ds^2 & = & -(N\,dt)^2 + h_{ij} (dx^i + V^i dt)(dx^j + V^j dt)
\ ,\eqnum{9a}\\
\bbox{A} & = & -\Phi dt + A_i (dx^i + V^i dt) \ .\eqnum{9b}
\eea
Here, $N$ is the lapse function, $V^i$ is the shift vector,
$h_{ij}$ is the spatial metric, $\Phi = -A_t + V^i A_i$ is the
scalar potential, and $A_i$ is the vector potential. The canonical
coordinates are $h_{ij}$ and $A_i$, and the canonically conjugate
momenta are
\bea
P^{ij} & = & -\frac{\sqrt{h}}{32\pi N} (h^{ij} h^{k\ell} -
h^{ik} h^{j\ell}) ({\dot h}_{k\ell} -
2 D_{{\sss (}k} V_{\ell{\sss )}}) \ ,\eqnum{10a}\\
\E^i & = & \frac{\sqrt{h}}{4\pi N} h^{ij} \Bigl( {\dot A}_j +
\partial_j(\Phi - V^k A_k) +
2 V^k \partial_{{\sss [}j} A_{k{\sss ]}}\Bigr) \ .\eqnum{10b}
\eea
In Eq.~(10a), $D_k$ denotes the covariant derivative in space.
The lapse $N$, shift $V^i$,
and scalar potential $\Phi$ appear in the Hamiltonian formalism
as Lagrange multipliers for the Hamiltonian, momentum, and
Gauss's law constraints, respectively.

Let \{${\tilde g}_\mn$, ${\tilde A}_\mu$\} denote any stationary
real Lorentzian solution of the Einstein--Maxwell equations,
written in stationary coordinates. In terms of the space--time
split (9), this solution is
\{$N$, $V$, $h$, $\Phi$, $A$\} =
\{$\tilde N$, ${\tilde V}$, ${\tilde h}$, ${\tilde\Phi}$,
${\tilde A}$\}, where the fields $\tilde N$, ${\tilde V}^i$,
${\tilde h}_{ij}$, ${\tilde\Phi}$, and ${\tilde A}_i$ are independent
of $t$ and are real. From this Lorentzian solution ({\it i.e.\/},
Eq.~(9) with tildes placed over the fields), the substitution
$t\to -it$ generates another field configuration, namely,
\{$N$, $V$, $h$, $\Phi$, $A$\} =
\{$\bar N$, ${\bar V}$, ${\bar h}$, ${\bar\Phi}$, ${\bar A}$\}, where
\bea
{\bar N} & = & -i {\tilde N} \ ,\qquad
{\bar V}^i = -i {\tilde V}^i \ , \qquad
{\bar\Phi} = -i {\tilde\Phi} \ ,  \eqnum{11a}\\
{\bar h}_{ij} & = & {\tilde h}_{ij} \ ,\qquad
{\bar A}_i = {\tilde A}_i \ .\eqnum{11b}
\eea
This is the instanton. Note that the lapse ${\bar N}$, shift
${\bar V}^i$, and scalar potential ${\bar\Phi}$ are imaginary.
If ${\bar V}^i = 0$, the metric for the instanton is real Euclidean;
otherwise the metric is complex.

The instanton is a solution of the Einstein--Maxwell
equations.\footnote{It serves no purpose to introduce the terminology
``Lorentzian equations of motion" and ``Euclidean equations of motion",
since a stationary Lorentzian solution and its associated instanton
satisfy the {\it same\/} equations of motion.} In the
Hamiltonian setting this follows from a few simple observations.
First, according to Eq.~(11b),
the canonical coordinates for the Lorentzian solution and the instanton
coincide. Also, definition (10) shows that the canonical momenta for
the Lorentzian solution are equal to the canonical momenta for the
instanton solution, ${\tilde P}^{ij} = {\bar P}^{ij}$ and
${\tilde\E}^i = {\bar\E}^i$. Thus, under the
substitution $t\to -it$, the canonical variables are unchanged and the
Lagrange multipliers are multiplied by the factor $-i$. Now, the
Einstein--Maxwell equations include the Hamiltonian, momentum,
and Gauss's law constraints. The constraints are constructed entirely
from the canonical variables---since they are satisfied for the
Lorentzian solution they are also satisfied for the instanton. The
remaining equations of motion are the evolution equations
${\dot f} = \{f,H\}$. Here, the brackets are Poisson brackets, $f$
denotes any function of the canonical variables, and the Hamiltonian
$H$ is a linear combination of constraints with Lagrange
multipliers as coefficients (plus suitable boundary terms).
For both the Lorentzian solution and
the instanton, the left--hand side ${\dot f}$ vanishes by stationarity.
Then for the Lorentzian solution
the right--hand side $\{f,H\}$ vanishes. The right--hand
side $\{f,H\}$ must vanish for the instanton case as well, since
it just differs from the right--hand side in the Lorentzian case by
an overall factor of $-i$. This shows that the equations of motion
${\dot f} = \{f,H\}$ are satisfied for the instanton configuration.

The stationary Lorentzian solution
\{$N$, $V$, $h$, $\Phi$, $A$\} =
\{$\tilde N$, ${\tilde V}$, ${\tilde h}$, ${\tilde\Phi}$,
${\tilde A}$\} and its associated instanton
\{$N$, $V$, $h$, $\Phi$, $A$\} =
\{$\bar N$, ${\bar V}$, ${\bar h}$, ${\bar\Phi}$,
${\bar A}$\} are characterized by the same canonical data,
including the electric field $E^i = -4\pi \E^i/\sqrt{h}$.
This is an essential feature of the instanton analysis. It
insures that the Lorentzian and instanton solutions match
along a stationary surface.
Also note that the value of the proper electrostatic potential
as determined by an (Eulerian) observer who is at rest in the
$t={\rm const}$ hypersurfaces, $-A_\mu U^\mu = \Phi/N$, is the
same for the Lorentzian and instanton solutions. Likewise, the
proper velocity of the spatial coordinate system, $V^i/N$, is
the same for the Lorentzian and instanton solutions. In certain
contexts this quantity has a physical meaning. For example,
for the thermodynamical description of a rotating black hole
\cite{BMY,BY} in corotating coordinates, $V^\phi/N$ is the
angular velocity of the black hole with respect to the
Eulerian observers.

The electric Ernst instanton solution is
\bea
{\bar g}_{\mn}dx^\mu dx^\nu & = & \frac{-\Lambda^2}{A^2(x-y)^2}
\Bigl[ G(y) dt^2 + G^{-1}(y) dy^2 - G^{-1}(x) dx^2 \Bigr]
+ \frac{G(x)}{A^2(x-y)^2\Lambda^2} d\varphi^2 \ ,\eqnum{12a}\\
{\EAbar}_t & = & \frac{iB}{2A^2} \biggl[ \frac{G(y)}{(x-y)^2}
(1 + Bqx - Bqy/2) + (1+r_- Ay)(1+r_+ Ay) (1-Bqy/2)\biggr]
\nonumber\\
& & -iqy -ik \ .\eqnum{12b}
\eea
The metric (12a) is real Euclidean since the shift vector for
the Ernst solution vanishes. The metric is regular for
$\xi_2 \leq y \leq \xi_3$ if $G'(\xi_2) = -G'(\xi_3)$ and if the
time coordinate is periodic with period
$4\pi/G'(\xi_3)$ \cite{GGS,GS,DGKT,Gibbons,Ross,DGGH,Yi,HHR}.
The vector field $\EAbar_\mu$ is
regular if $\EAbar_t$ vanishes at both $y=\xi_2$ and $y=\xi_3$.
These conditions are satisfied if $\EAbar_\mu$ is defined
separately in open neighborhoods of $y=\xi_2$ and $y=\xi_3$,
and in each of these neighborhoods the constant $k$ of
Eq.~(12b) is chosen appropriately.

Topologically, the Ernst instanton can be viewed as $\RR^4$
with the interior of a ``tube" $S^1\times S^2$ removed, and
points along the $S^1$ direction identified. This
two--dimensional surface is $y=\xi_2$, and is referred to below
as the Euclidean wormhole. The acceleration horizon of the
Lorentzian Ernst solution corresponds to the two--dimensional
surface $y = \xi_3$ of the instanton solution. For the instanton,
the wormhole $y=\xi_2$ surrounds the surface $y=\xi_3$.

%%%%%%%%%%%%%%%%%%%%%%%%%%%%%%%%%%%%%%%%%%%%%%%%%%%%%%%%%%%%%%%%%%%%
%%%%%%%%%%%%%%%%%%%%    CHAPTER FOUR   %%%%%%%%%%%%%%%%%%%%%%%%%%%%%
%%%%%%%%%%%%%%%%%%%%%%%%%%%%%%%%%%%%%%%%%%%%%%%%%%%%%%%%%%%%%%%%%%%%
\section{Black Hole Pair Creation in Einstein--Maxwell Theory}
In the path integral for Einstein--Maxwell theory, each history
\{$g_\mn$, $A_\mu$\} enters with a weight $\exp(\S)$ where
\be
\S[g_\mn,A_\mu] = \frac{i}{16\pi} \int_{\M} d^4x  \sqrt{-g}
(R -F_\mn F^\mn) + (\hbox{boundary terms}) \ .\eqnum{13}
\ee
I will refer to $\S$ as the action.
The path integral is ultimately defined as a
sum over either Lorentzian metrics, or Euclidean metrics, or some
other class of metrics. For the purpose of computing the leading
order (exponential) contribution to the path integral this issue is
not important. In particular, the instanton can be viewed as a stationary
point in a sum over real $N$, $V$, $h$, $\Phi$, and $A$
that lies off the axis of integration. Alternatively, one  can
rotate the integration contours for $N$, $V$, and $\Phi$ in the
complex plane so that the instanton lies on the axis of integration.

The boundary terms in $\S$ depend on the boundary conditions that are
appropriate for the problem at hand. Here, the pair creation rate for
electrically charged black holes is computed relative to the pair
creation rate for electrically charged stars. In the
instanton approximation this is given by the exponential of $\S$ for the
electric Ernst instanton (eEi) divided by the exponential of $\S$ for
the charged star instanton (csi). Thus, all that is required is the
difference $\S[{\rm eEi}] - \S[{\rm csi}]$. I will assume that the
stars are compact, and that the matter, charge, and stress inside
the stars are distributed in such a way that the gravitational and
electromagnetic fields outside the stars coincide with the fields of
the electric Ernst solution exterior to a pair of closed surfaces that
surround the black holes. Then the instanton for the charged star is
topologically $\RR^4$ and coincides with the electric Ernst instanton
everywhere except in the interior of a ``tube" $S^1\times S^2$ that
encompasses the matter (for the charged star instanton) or wormhole
(for the Ernst instanton).
In this case the boundary terms that appear in Eq.~(13) cancel in the
calculation of $\S[{\rm eEi}] - \S[{\rm csi}]$.

The calculation $\S[{\rm eEi}] - \S[{\rm csi}]$ is easily carried
out using the Hamiltonian approach with the electric Ernst and
charged star instantons foliated along the surfaces of constant
stationary time $t$. The Hamiltonian form of the action for
Einstein--Maxwell theory is
\be
\S = i \int dt\, d^3x \Bigl( P^{ij} {\dot h}_{ij} + \E^i {\dot A}_i
- N\H - V^i\H_i + \Phi\G \Bigr) + (\hbox{boundary terms})
\ ,\eqnum{14}
\ee
where $\H$, $\H_i$, and $\G$ are the Hamiltonian, momentum, and
Gauss's law constraints. For the theory that describes the charged
star solution, the matter fields contribute extra ``$p\dot q$" terms
to $\S$ and also contribute terms to the constraints. (The matter
fields might also contribute boundary terms at infinity. These will
vanish for the charged star instanton since the matter distribution
is compact.) The boundary terms in Eq.~(14) include the boundary
terms at infinity from Eq.~(13) and also boundary terms that
arise from total derivatives and integrations by parts in the
space--time decomposition.\footnote{For Einstein gravity, a
systematic derivation of the Hamiltonian action (14) from the
covariant action (13), including boundary terms, was given in
Ref.~\cite{BY}. Electromagnetic, Yang--Mills, and other matter
fields can be incorporated into that derivation in a
straightforward manner.} The surfaces  $t={\rm const}$
extend from infinity to the ``acceleration horizon"
two--surface $y=\xi_3$, which serves as a boundary for the
three--dimensional hypersurfaces. The boundary terms in Eq.~(14)
include boundary terms at $y=\xi_3$. These terms, like the boundary
terms at infinity, cancel in the calculation of the difference
$\S[{\rm eEi}] - \S[{\rm csi}]$.

For the electric  Ernst instanton, but not for the charged star
instanton, the hypersurfaces $t={\rm const}$ intersect at the
Euclidean wormhole. This two--dimensional surface constitutes
an inner boundary $b$ of topology $S^2$ for the
hypersurfaces. In passing from the Lagrangian form (13)
to the Hamiltonian form (14) of $\S$, the various total derivatives
and integrations by parts introduce boundary terms at $b$. These
boundary terms can be derived by cutting out a small region
surrounding the wormhole, then taking the limit as the excised
region vanishes. In this way an inner boundary $B$ of topology
$S^2\times S^1$ is introduced into the Ernst instanton manifold.
Under the simplifying assumption that the (outward pointing) unit
normal vector field $n^\mu$ of $B$ lies in the $t={\rm const}$
hypersurfaces, the boundary terms are \cite{BMY,BY,CECS}
\be
\S\bigr|_{B} = -i\int dt\int_{b} d^2x \sqrt{\sigma}
\Bigl[ (n^i \partial_i N)/(8\pi) + V^i n^j(2P_{ij} +
A_i\E_j)/\sqrt{h}  - \Phi n_i\E^i/\sqrt{h} \Bigr] \ .\eqnum{15}
\ee
Here, $\sigma$ denotes the determinant of the metric on $b$.

With $\S$ written in Hamiltonian form (14), all boundary terms
except those displayed in Eq.~(15) cancel in the calculation
$\S[{\rm eEi}] - \S[{\rm csi}]$. Furthermore, for both the
electric Ernst instanton and the charged star instanton, the
(four--dimensional) volume integral terms in $\S$ vanish---the
``$p\dot q$" terms vanish by stationarity and the remaining
terms vanish because the constraints hold. Therefore
$\S[{\rm eEi}] - \S[{\rm csi}]$ is equal to the boundary term
(15) evaluated at the electric Ernst instanton. Recall that for
the Ernst instanton the shift vector $V^i = {\bar V}^i$ is zero
and by regularity the scalar potential
$\Phi = -\EAbar_t$ vanishes at the wormhole $b$. Thus
only the first term in Eq.~(15) is nonzero. Setting
$N={\bar N} = -i{\tilde N}$, we have
\be
\S[{\rm eEi}] - \S[{\rm csi}] =
-\int dt\int_{b} d^2x \sqrt{\sigma}
(n^i\partial_i {\tilde N})/(8\pi) \ .\eqnum{16}
\ee
At each point of $b$, the quantity $-\int dt(n^i\partial_i
{\tilde N})$ is the rate of change of proper circumference
with respect to proper radius for the circular trajectories
of $\partial/\partial t$ in the neighborhood of $B$.
(The minus sign appears because the
normal $n^\mu$ points in the direction of decreasing radius.)
By regularity of the metric this equals $2\pi$. Therefore
Eq.~(16) becomes
\be
\S[{\rm eEi}] - \S[{\rm csi}] = A_{\sss BH}/4 \ ,\eqnum{17}
\ee
where $A_{\sss BH}$ is the area $\int_{b} d^2x \sqrt{\sigma}$
of the wormhole. In turn, $A_{\sss BH}$ equals the horizon
area of each black hole in the Lorentzian Ernst solution.

Equation (17) shows that, in the instanton approximation, the
pair creation rate for electrically charged black holes is
enhanced by a factor of $\exp(A_{\sss BH}/4)$ relative to the
pair creation rate for electrically charged stars.
Note that, with the calculation organized as
above, the detailed forms of the electric Ernst solution and the
charged star solution are not needed. Thus, the result (17)
shows that the pair creation rate for black holes in
Einstein--Maxwell theory is always enhanced by the factor
$\exp(A_{\sss BH}/4)$ relative to the pair creation rate for
matter distributions.

%%%%%%%%%%%%%%%%%%%%%%%%%%%%%%%%%%%%%%%%%%%%%%%%%%%%%%%%%%%%%%%%%%%%
%%%%%%%%%%%%%%%%%%%%    CHAPTER FIVE   %%%%%%%%%%%%%%%%%%%%%%%%%%%%%
%%%%%%%%%%%%%%%%%%%%%%%%%%%%%%%%%%%%%%%%%%%%%%%%%%%%%%%%%%%%%%%%%%%%
\section{Black Hole Pair Creation in General}
The inner boundary terms (15) that yield the  black hole entropy factor
for pair creation in Einstein--Maxwell theory are precisely the same
terms that yield the black hole entropy in the path integral analysis
of the partition functions \cite{BMY,BY,CECS}. The partition functions
are obtained from the density of states $\nu$ by Laplace transforms---for
example, the grand canonical partition function is obtained from
$\nu$ by a Laplace transform that replaces energy with inverse temperature
as the independent thermodynamical variable.
For the purpose of deriving the entropy it is most convenient to work
directly with the density of states $\nu$.  The entropy is given
by the logarithm of $\nu$.

The density of states is a function of the
thermodynamical extensive variables such as energy, angular momentum,
electric charge, {\it etc\/}. Expressed as a path integral, $\nu$ is
a sum over all fields that fit inside an outer boundary (the
periodically identified history of a ``box")  of topology
$S^2\times S^1$. The extensive variables are fixed as boundary
conditions on this outer boundary \cite{BY}. Consider the action for
such a path integral, for any generally covariant theory of
gravitational and matter fields. For the moment, let the manifold
$\M$ have topology $\M=\Sigma\times S^1$, where space $\Sigma$ has
boundary $\partial\Sigma = S^2$ and $\M$ has boundary
$\partial\M = S^2\times S^1$. I will assume that the action
can be written in Hamiltonian form
\be
\S[\lambda,q,p] = i \int dt\, d^3x \Bigl(  p_a {\dot q}^a
- \lambda^{\sss A}{\cal C}_{\sss A} \Bigr) + ({\hbox{boundary
terms}}) \ .\eqnum{18}
\ee
Notice that the volume integral contribution to the
Hamiltonian is written as a linear combination of constraints
${\cal C}_{\sss A}({q},{p})$ with Lagrange multipliers $\lambda^{\sss A}$.
This form for the Hamiltonian follows from general covariance and the
property that under reparametrizations in $t$ the canonical
variables $ p_a$ and $ q^a$ transform as scalars and the Lagrange
multipliers transform as scalar densities \cite{HandT}. I will assume
that these conditions hold. It also follows that the boundary terms
in (18) cannot depend solely on the canonical
variables---each term, expressed as an integral over $\partial\M$, must
include a Lagrange multiplier as a factor in its integrand in order to
transform properly under reparametrizations in $t$.

The boundary terms in the action (18) must
be correlated with the boundary conditions on $\partial\M$ in such a
way that the boundary terms in the
variation $\delta\S$ vanish. There are two types of boundary terms
in $\delta\S$, namely, those that arise from variation of the boundary terms
in $\S$, and those that arise from integration by parts.
Integration by
parts occurs when the constraints ${\cal C}_{\sss A}$ contain spatial
derivatives
of the canonical variables. Thus, the boundary terms in $\delta\S$ that
arise through integration by parts necessarily involve variations of the
canonical variables. On the other hand, the explicit boundary terms in $\S$,
upon variation, generate boundary terms in $\delta\S$ that involve
variations of the Lagrange multipliers. These boundary terms
can never be canceled by the boundary terms that come from integration by
parts.
Consequently, if the action $\S$ includes
any explicit boundary terms, then $\delta\S$ will include boundary
terms that involve variations of quantities that depend on the Lagrange
multipliers.
Because the
boundary terms in $\delta\S$ must vanish by virtue of the boundary
conditions,
we see that the boundary data for this action will include fixation
of quantities that depend on the Lagrange multipliers.

Armed with these observations it follows that the action appropriate
for the path integral representation of the density of states $\nu$
(the ``microcanonical action" \cite{BY}) is given by Eq.~(18) with
{\it no boundary terms\/}. Here is the reason:
The density of states is a function of the thermodynamical extensive
variables which, by definition, are properties of the
states of the system. These variables appear at the classical level as
functions of the canonical variables $p_a$ and $q^a$. Thus, the path
integral for $\nu$ must come from an action in which the fixed boundary
data are functions only of the canonical variables---by the arguments above,
such an action has no explicit boundary terms.

Now consider a stationary Lorentzian black hole solution
\{${\tilde\lambda}^{\sss A}$,
${\tilde q}^a$, ${\tilde p}_a$\} of the classical equations
of motion. The black hole's entropy is
found by evaluating the path integral for the density
of states $\nu$, with the boundary data fixed to those values that
characterize
the black hole. The path integral is given approximately by its integrand
$\exp(\S[\lambda,q,p])$ evaluated at a complex black hole extremum
\{${\bar\lambda}^{\sss A}$, ${\bar q}^a$, ${\bar p}_a$\}.
The complex black hole is obtained from the Lorentzian black hole
by the relations
\be
{\bar\lambda}^{\sss A} = -i{\tilde\lambda}^{\sss A} \ ,\qquad
{\bar q}^a = {\tilde q}^a \ ,\qquad
{\bar p}_a = {\tilde p}_a \ .\eqnum{19}
\ee
The complex black hole satisfies the boundary conditions, since the
boundary conditions involve only the canonical variables.
The fact that the complex black hole is a solution of the classical equations
of motion follows
from an obvious generalization of the arguments given in Sec.~3.
Alternatively, observe that
\{${\bar\lambda}^{\sss A}$, ${\bar q}^a$, ${\bar p}_a$\} is obtained from
\{${\tilde\lambda}^{\sss A}$, ${\tilde q}^a$, ${\tilde p}_a$\}
by a reparametrization $t\to -it$, where $p_a$ and $q^a$ transform as
scalars and $\lambda^{\sss A}$ transforms as a scalar density.

The path integral constructed from the action (18) with no boundary
terms yields the contribution to $\nu$ from the topological sector
$\Sigma\times S^1$. However, the complex black hole is an extremum
of the action on a manifold with topology
$\M = S^2\times\RR^2$ and boundary $\partial\M = S^2\times S^1$. Thus,
the action
(18) with no boundary terms is not correct as it stands for use in the
approximation $\exp(\S[{\bar\lambda},{\bar q},{\bar p}])$ to the
density of states. It is, however, correct with regard to
the lack of boundary terms at $\partial\M$. The action to be
used in the approximate evaluation of $\nu$ can be found by starting from
(18), with no boundary terms, and writing this action in manifestly
covariant (Lagrangian) form. The manifold can then be
chosen to have topology $S^2\times\RR^2$. The resulting Lagrangian action
can be written in Hamiltonian form
if a small region surrounding the center of the disk $\RR^2$, where the
hypersurfaces intersect, is removed. This introduces an inner boundary
$B=S^2\times S^1$ in $\M$. In passing to the Hamiltonian form of the
action, total derivatives and integrations by parts will generate
boundary terms at the inner boundary $B$.
There will be no boundary terms at the outer boundary, however, since
none were present in the original action.

{}From the discussions above it follows that the black hole entropy is
approximated by $\S_{\sss BH} \approx
\S[{\bar\lambda},{\bar q},{\bar p}]$, and
the action $\S[\lambda,q,p]$ has the
form (18) where {\it only inner boundary terms are present\/}.
In evaluating this action at the complex solution
\{${\bar\lambda}^{\sss A}$, ${\bar q}^a$, ${\bar p}_a$\},
the limit is taken in which the excised region vanishes.
Since $\partial{\bar q}/\partial t=0$ (by stationarity) and
${\cal C}_{\sss A}(\bar q,\bar p) = 0$, the only nonzero contribution to
the entropy $\S_{\sss BH} \approx \S[{\bar\lambda},{\bar q},{\bar p}]$
comes from the inner boundary terms.\footnote{It should be emphasized
that the method discussed here, which generalizes the analysis of
Refs.~\cite{BMY,BY,CECS}, can be used to derive an explicit expression
for black hole entropy for any generally covariant theory of
gravitational and matter fields that can be placed in Hamiltonian form.
This method shows that the entropy $\S_{\sss BH} \approx
\S[{\bar\lambda},{\bar q},{\bar p}]$ is a ``geometrical" quantity
(constructed from the gravitational and matter fields)
defined locally at the black hole horizon. The local, geometrical
character of black hole entropy has been examined in detail in
Ref.~\cite{Wald} using Noether charge techniques.} Note that the
resulting entropy is real, since each inner boundary term must
contain a Lagrange multiplier factor (in order to transform properly
under reparametrizations) and the Lagrange multipliers
${\bar\lambda}^{\sss A}$ are imaginary.

The inner boundary terms that yield the black hole entropy coincide with
the inner boundary terms that yield the relative enhancement
factor for black hole pair creation. This is not
difficult to see. Recall from the example of black hole pair creation in
Einstein--Maxwell theory that the action for the instanton that describes
black hole pair creation, when written
in Hamiltonian form, includes  boundary terms at infinity as
well as boundary terms at the acceleration and black hole horizons.
The action for the instanton that describes pair creation of matter
distributions (stars) contains the same boundary terms at infinity and
at the acceleration horizon, but of course no horizon boundary terms.
The volume integral contributions to the Hamiltonian actions for both the
black hole instanton and the star instanton vanish because the instantons
are stationary and satisfy the constraints. Thus, in taking the difference
between the
actions for the black hole instanton and the star instanton, only the
inner boundary terms from the black hole horizon survive. Those inner
boundary terms are derived by the same analysis as the inner boundary
terms for black hole entropy. Namely,
they arise when the Lagrangian action is expressed in Hamiltonian
form in the presence of the boundary $B$ of a small excised region
around the black hole event horizon where the hypersurfaces
intersect.

The enhancement factor for black hole pair creation is obtained
by evaluating the inner boundary terms at the black hole instanton
solution, while the entropy of a black hole is obtained by evaluating
the {\it same\/} inner boundary terms at the complex black hole
solution. But the black hole instanton is related to a real Lorentzian
solution, which represents a physical black hole pair, by the
substitution $t\to -it$. This relationship agrees precisely
with the relationship between either of the two physical black holes and
the complex solution that yields its entropy. We are therefore led to the
main conclusion that the enhancement in the pair creation rate for black
holes is given by the factor $\exp(\S_{\sss BH})$ for any
generally covariant theory of gravitational and matter fields.

%%%%%%%%%%%%%%%%%%%%%%%%%%%%%%%%%%%%%%%%%%%%%%%%%%%%%%%%%%%%%%%%%%%%%%%%
%%%%%%%%%%%%%%%%%%%%    ACKNOWLEDGMENTS    %%%%%%%%%%%%%%%%%%%%%%%%%%%%%
%%%%%%%%%%%%%%%%%%%%%%%%%%%%%%%%%%%%%%%%%%%%%%%%%%%%%%%%%%%%%%%%%%%%%%%%
\section{Acknowledgments}
I would like to thank G.~T.~Horowitz and R.~M.~Wald for helpful remarks,
and J.~W.~York for helpful discussions and comments on the manuscript.

%%%%%%%%%%%%%%%%%%%%%%%%%%%%%%%%%%%%%%%%%%%%%%%%%%%%%%%%%%%%%%%%%%%%%%%%
%%%%%%%%%%%%%%%%%      REFERENCES       %%%%%%%%%%%%%%%%%%%%%%%%%%%%%%%%
%%%%%%%%%%%%%%%%%%%%%%%%%%%%%%%%%%%%%%%%%%%%%%%%%%%%%%%%%%%%%%%%%%%%%%%%

\end{document}